\documentclass[11pt]{article}
\usepackage{moriond,epsfig}

\def\be{\begin{equation}}
\def\ee{\end{equation}}
\def\bea{\begin{eqnarray}}
\def\eea{\end{eqnarray}}

\newcommand{\MET}{\mbox{$\raisebox{.3ex}{$\not\!$}E_T$}}
\newcommand{\METVEC}{\mbox{$\raisebox{.3ex}{$\not\!$}{\vec E}_T$}}

\begin{document}

%\begin{flushright}
%CDF/PUB/TOP/PUBLIC/9770\\
%\today; ~v3.0
%\end{flushright}

\vspace*{4cm}
\title{OBSERVATION OF ELECTROWEAK SINGLE TOP QUARK PRODUCTION AT THE TEVATRON}

\author{ENRIQUE PALENCIA}

\address{Fermi National Accelerator Laboratory, Batavia, Illinois 60510, USA}

%\address{Department of Physics, Theoretical Physics, 1 Keble Road,\\
%Oxford OX1 3NP, England}

\maketitle\abstracts{
We report the first observation of single top quark production using $p{\bar{p}}$ collision data 
with $\sqrt{s}=1.96$~TeV collected by the CDF II and D0 detectors at Fermilab.  The significance of both the 
observed CDF and D0 data is 5.0 standard deviations, and the expected sensitivity is in excess of 5.9 
and equal to 4.5 standard deviations, respectively. The single top production cross section and the CKM matrix 
element value \mbox{$|V_{tb}|$} have been measured.
%We measure a cross section of $2.3^{+0.6}_{-0.5}(\mathrm{stat+syst})$~pb, extract the CKM matrix 
%element value \mbox{$|V_{tb}|=0.91^{+0.11}_{-0.11} (\mathrm{stat+syst})\pm 0.07(\mathrm{theory})$}, 
%and set the limit $|V_{tb}|>0.71$ at the 95\% C.L.
}

\section{Introduction}
The establishment of the presence of the electroweak production of single top quarks in $p\bar{p}$ 
collisions is an important goal of the Tevatron program.  The reasons for studying single top quarks 
are compelling: the production cross section is directly proportional to the square of the CKM 
matrix~\cite{ckm} element $|V_{tb}|$, and thus a measurement of the rate constrains fourth-generation 
models, models with flavor-changing neutral currents, and other new phenomena~\cite{Tait:2000sh}. 
Furthermore, %because single top quark production is a well-understood process in the standard model (SM), 
understanding single top quark production provides a solid anchor to test the analysis techniques that 
are also used to search for Higgs boson production and other more speculative phenomena.

In the SM, top quarks are expected to be produced singly through $t$-channel or $s$-channel exchange 
of a virtual $W$ boson. %as shown in Fig.~\ref{fig:sttfeyn}. 
This electroweak production of single top 
quarks is a really difficult process to measure because the expected combined production cross section 
($\sigma_{s+t}\sim2.9$~pb~\cite{harris,stxs}) is much smaller than those of competing background processes. 
Also, the presence of only one top quark in the event provides fewer features to use in separating the 
signal from background, compared with measurements of top pair production ($t{\bar{t}}$), which was first 
observed in 1995~\cite{ttbar}.

%\rule{5cm}{0.2mm}\hfill\rule{5cm}{0.2mm}
%\vskip 2.5cm
%\rule{5cm}{0.2mm}\hfill\rule{5cm}{0.2mm}
%\begin{figure}
%\psfig{figure=figures/sttfeyn.eps,height=2.7in}
%\vskip -2.cm
%\caption{Representative Feynman diagrams of single top quark production. Figures (a) and (b)
%are $t$-channel processes, and Fig. (c) is the $s$-channel process.
%\label{fig:sttfeyn}}
%\end{figure}

Both the CDF and D0 collaborations have published evidence for single top quark production at 
significance levels of 3.7 and 3.6 standard deviations, respectively~\cite{CDF,d0evidence}. This 
article describes the latest analysis done using data collected with the CDF II~\cite{CDFdet} (with 
3.2 fb$^{-1}$) and D0~\cite{D0det} (with 2.3 fb$^{-1}$) detectors at the Tevatron and reports 
observation of single top quark production~\cite{cdfObsPRL,d0ObsPRL}.

Since the two collaborations use similar analysis techniques, the next sections apply to both of %them unless other thing is stated. 
the analyses unless otherwise is stated.

\section{Event Selection and Backgrounds}
For the analyses shown here, we assume that single top quarks are produced in the
$s$- and $t$-channel modes with the SM ratio, and that the branching fraction of the top quark
to $Wb$ is 100\% (corresponding to $|V_{tb}|~>>~|V_{ts}|,~|V_{td}|$). For most of the analysis channels, 
we seek events in which the $W$ boson decays leptonically in order to improve the signal-to-background ratio 
$s/b$. %We simulate single top events using a tree-level matrix-element generator~??? as in Ref.~???. 

%Three distinct trigger algorithms are employed to select the data used in this analysis: a 
%high pT electron trigger, a high pT muon trigger, and a trigger that requires large missing 
%transverse energy with either an energetic electromagnetic cluster or two separated jets [7,8]???.

The basic event selection is based on selecting $\ell$+\protect\MET+jets events, where
$\ell$ is an explicitly reconstructed electron or muon from the $W$ boson decay. 
This lepton is required to be isolated from nearby jets and to have large transverse momentum. The 
presence of high missing transverse energy (\MET) and at least two energetic jets
is also required. At least one of the jets has to be identified as containing a $B$ hadron.

The background has contributions from events in which a $W$ boson is produced in 
association with one or more heavy flavor jets, events with mistakenly 
$b$-tagged light-flavor jets, multijet events (QCD), $t{\bar{t}}$ and 
diboson processes, as well as $Z$+jet events. %The expected event yields in Table~??? 
%are estimated as in Ref.~??? where the signal, $t{\bar{t}}$, and diboson categories 
%are Monte Carlo predictions scaled to the total integrated luminosity while the 
%remaining categories use predictions derived from control samples taken from the full 
%event sample.

\begin{figure}[h]
\begin{minipage}{0.5\linewidth}
\begin{center}
\psfig{figure=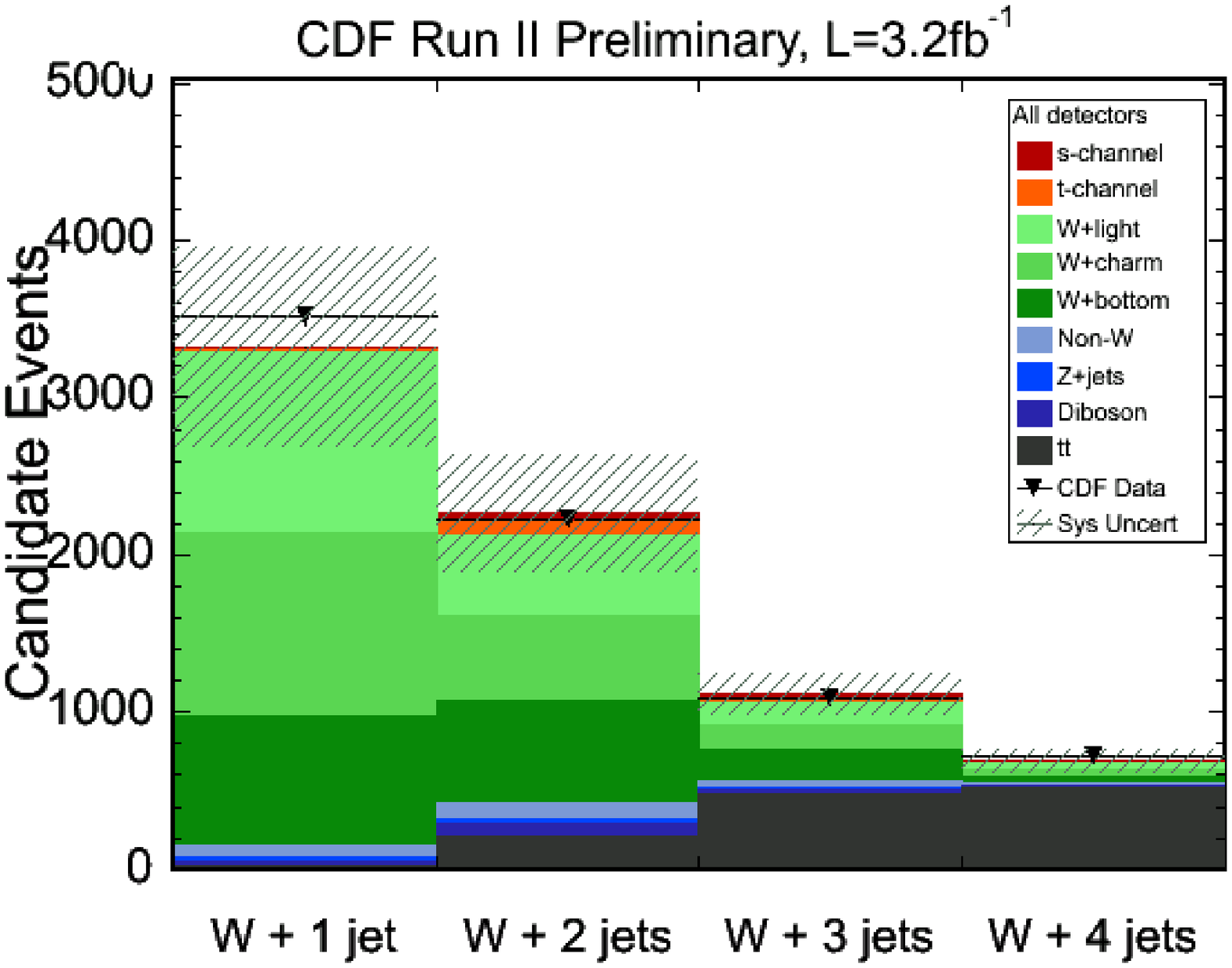,width=.92\textwidth}
%\caption{Expected number of events as a function of the number of jets for the signal and each
%background process.
%\label{fig:yieldsfig}}
\end{center}
\end{minipage} %\hspace{.2cm}
\begin{minipage}{0.5\linewidth}
\begin{center}
   \psfig{figure=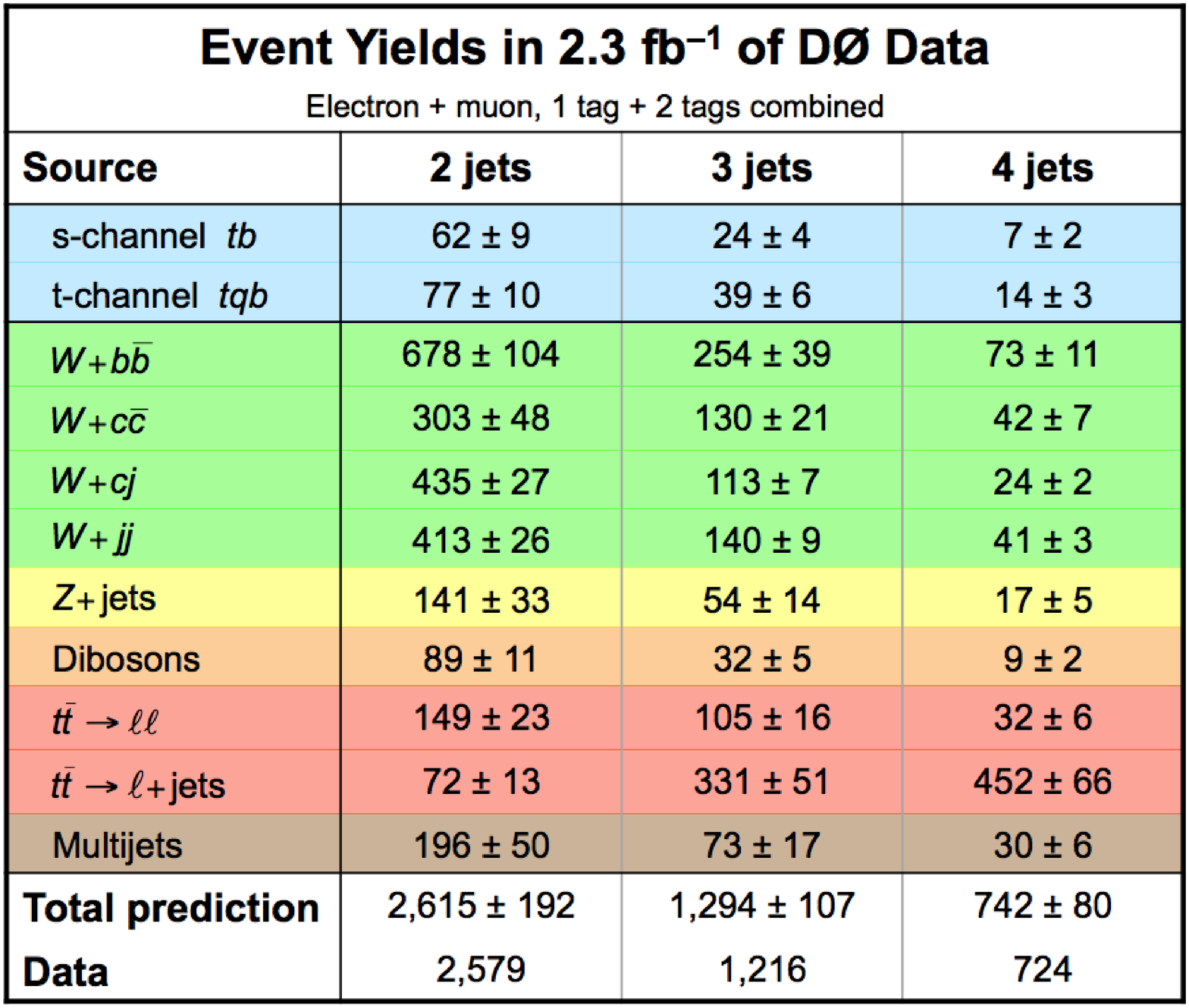,width=.8\textwidth}
%\caption{D0 yields for events with 2, 3 and 4 jets.
%\label{fig:D0yields}}
\end{center}
\end{minipage} 
\caption{Left: Expected number of CDF $\ell$+\protect\MET+jets events as a function of the number of jets for the signal and each
background process. Right: D0 yields for events with 2, 3 and 4 jets.
\label{fig:yields}}
\end{figure}

The expected number of $\ell$+\protect\MET+jets events, in CDF, as a function of the number of jets for the signal and each
background process is shown in Fig.~\ref{fig:yields} (left). The D0 yields for events with 2, 3 or 4 jets
are shown in Fig.~\ref{fig:yields} (right). From these figures, it is clear that
single top signal is hidden under huge and uncertain backgorunds which make counting 
experiments impossible. %Instead, multivariate analyses are needed to discriminate single top from the 
%backgrounds.

%\rule{5cm}{0.2mm}\hfill\rule{5cm}{0.2mm}
%\vskip 2.5cm
%\rule{5cm}{0.2mm}\hfill\rule{5cm}{0.2mm}

\section{Multivariate Analysis}
To overcome these challenges, a variety of multivariate techniques for separating single top 
events from the backgrounds have been developed as described below. 

\subsection{Likelihood Function (LF)}
This technique is used only by the CDF collaboration. A projective likelihood technique~\cite{lfcitation} 
is used to combine information from several input variables to optimize the separation of the single 
top signal from the backgrounds. Two likelihood functions are created, one for two jet events, and one for 
three jet events with 7 and 10 input variables, respectively. Some of the input variables 
used are: the total scalar sum of transverse energy in the event $H_T$, $Q \times \eta$~\cite{cpyuan}, the dijet mass 
$M_{jj}$, cos$\theta^*_{lj}$~\cite{costheta} and the t-channel matrix element.

A new separate search for the single top in the s-channel is also done using this technique (LFS). In this 
case, the likelihood function is optimized to be sensitive to the $s$-channel process. %(Fig.~\ref{fig:sttfeyn}(c)) 
using the subset of the $\ell$+\protect\MET+jets sample with two $b$-tagged jets.

\subsection{Neural Networks (NN)}
This approach employs neural networks which combine many variables into one more powerfull discriminant and 
have the general advantage that correlations between the discriminating input variables are identified and 
utilized to optimize the separation power between signal and background. D0 uses Bayesian NN (BNN), which average
over hundreds of networks for each analysis channel to obtain better separation.

%The networks used by CDF are developed using the NEUROBAYES analysis package [20]???, which combines a 
%three-layer feed-forward neural network with a complex and robust preprocessing of the input variables. 
%Bayesian regularization techniques are utilized to avoid overtraining.

%Four separate networks are trained to identify different signals in distinct samples using simulated events 
%from the common samples described previously. The networks use 11 to 18 input variables. The most important 
%ones are Mlnub, bnn, Mjj, Q times eta, cos(l,j), the transverse mass of the W boson, and HT. The input variables
%are selected from a large list using an automated evaluation during the preprocessing step before the network
%training. In an iterative process, we determine those variables whose removal would cause a significant loss in
%separation power between signal and background and use them for network training.

\subsection{Matrix Elements (ME)}
The ME method relies on the evaluation of event probability densities for signal and background 
processes based on calculations of the standard model differential cross sections~\cite{me}. We construct 
these probability densities for each signal and background process for each event given their measured quantities $x$ by integrating 
the appropriate differential cross section $d\sigma(y)/dy$ over the underlying partonic quantities $y$, convolved 
with the parton distribution functions (PDFs) and detector resolution effects.

The event probability densities are combined into an event probability discriminant: 
$EPD = P_{signal}/(P_{signal} + P_{background})$. To better classify signal events that contain $b$ jets, 
the CDF collaboration incorporates the output of a neural network jet-flavor separator~\cite{Richter:2007zzc} 
into the final discriminant. D0 applies the NN tagging probability to each jet and weights
all combinations appropriately.

\subsection{Boosted Decission Trees (BDT)}
The BDT discriminant uses a decision tree method that applies binary cuts iteratively to classify events~\cite{DT}. 
The discrimination is further improved using a boosting algorithm~\cite{TMVA}. %~\cite{AdaBoost,TMVA}. 
The BDT discriminant uses over 20 input variables in the case of CDF and 64 in D0. Some of the most sensitive are the neural-network jet-flavor 
separator (only in CDF), the invariant mass of the $\ell\nu b$ system $M_{\ell\nu b}$, the total scalar sum of transverse 
energy in the event $H_\mathrm{T}$, $Q\times\eta$, the dijet mass $M_{jj}$, and the transverse mass of the $W$ boson. 

\subsection{\protect\MET + jets (MJ)}
The MJ analysis is a new analysis in CDF designed to select events with \MET~and jets, while vetoing events selected by 
the $\ell+\MET$+jet analyses.  It accepts events in which the $W$ boson decays into $\tau$ leptons and those in which 
the electron or muon fails the lepton identification criteria. 

%We use data corresponding to 2.1 fb$^{-1}$ of integrated luminosity for the MJ analysis and select events
%that have $\MET > 50$~GeV and two jets within
%$|\eta|<2.0$, at least one of which has $|\eta|<0.9$.  
%The jet energy measurements include information from both the calorimeter and the 
%charged-particle spectrometer.  Events must have one jet with transverse energy
%$E_T$ greater than 35~GeV, and a second jet with $E_T$ greater than 25~GeV.  The angular separation between the two jets, $\Delta R=\sqrt{(\Delta\eta)^2+(\Delta\phi)^2}$, 
%is required to exceed 1.0.   We reject events with four or more jets
%with $E_T>15$~GeV in $|\eta|<2.4$ in order to reduce the multijet (QCD) and 
%$t{\bar{t}}$ backgrounds.  We identify $b$ jets with the same
%algorithm as Ref.~??? supplemented with a jet probability algorithm~???.  

The advantage of this analysis is that it is orthogonal to the $\ell$+\protect\MET+jets analysis 
described above, increasing the signal acceptance by $\sim$30\%. The disadvantage is the huge instrumental 
background due to QCD events in which mismeasured jet energies produce large~\METVEC\ aligned in 
the same direction as jets. To reduce this background, a neural network is used removing 77\% of 
the QCD background while keeping 91\% of the signal acceptance.

%We use the transverse momentum imbalance ($\MPTVEC$) as measured in the spectrometer. This variable 
%is more correlated to the neutrino energy and its direction than $\METVEC$ in this class of events. 
%The absolute amount of $\MET$ and mpt, the angle between them, the azimuthal angles between $\METVEC$ 
%or $\MPTVEC$ and the jet directions, and several other less powerful variables are used as inputs to 
%a neural network (NNQCD).  The NNQCD output is required to pass a threshold, removing 77\% of the QCD 
%background while keeping 91\% of the signal acceptance.

%The backgrounds in the MJ analysis due to QCD events and events with light flavor jets produced in 
%association with $W$ and $Z$ bosons are estimated using data in a control region composed of events 
%in which the $\METVEC$ is aligned with one of the jets.  The observed and expected event counts for 
%the MJ analysis are given in the \MET+jets column of Table~\ref{tab:events}.

Finally, the MJ discriminant uses a neural network to combine information from several input variables. The 
most important variables are the invariant mass of the $\METVEC$ and the second leading jet, the scalar sum
of the jet energies, the $\MET$, and the azimuthal angle between the $\METVEC$ and the jets.

\subsection{Combination}
D0 combines the  ME, BNN and BDT channels using a bayesian neural network. The three discriminant
outputs from each analysis channel are used as inputs to a Bayesian neural
network, obtaining a single discriminant output for each channel. As a cross-check, the
Best Linear Unbiased Estimator (BLUE)~\cite{blue} is used.

CDF combines the LF, ME, NN, BDT, and LFS channels using a super-discriminant (SD) technique. The SD method 
uses a neural network trained with neuro-evolution~\cite{neat} to separate the signal from the background 
taking as inputs the discriminant outputs of the five analyses for each event. %With the super-discriminant 
%analysis we improve the sensitivity (defined below) by 13\% over the best individual analysis. 
A simultaneous fit over the two exclusive channels, MJ and SD, is performed to obtain the final combined results
(see next Section). 

\begin{figure}[[h]
\begin{minipage}{0.5\linewidth}
   \psfig{figure=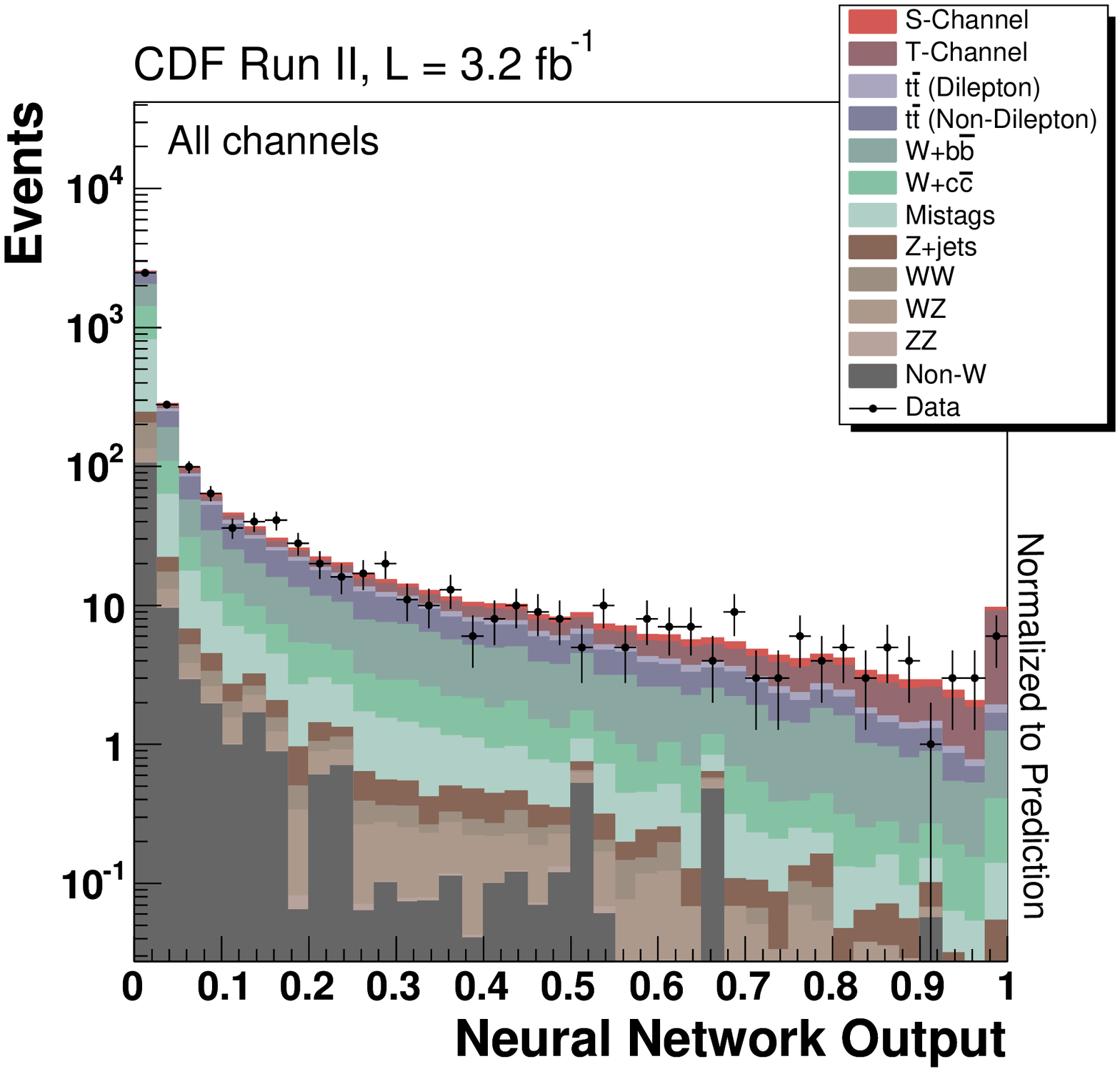,width=.98\textwidth}
\end{minipage} %\hspace{.2cm}
\begin{minipage}{0.5\linewidth}
   \psfig{figure=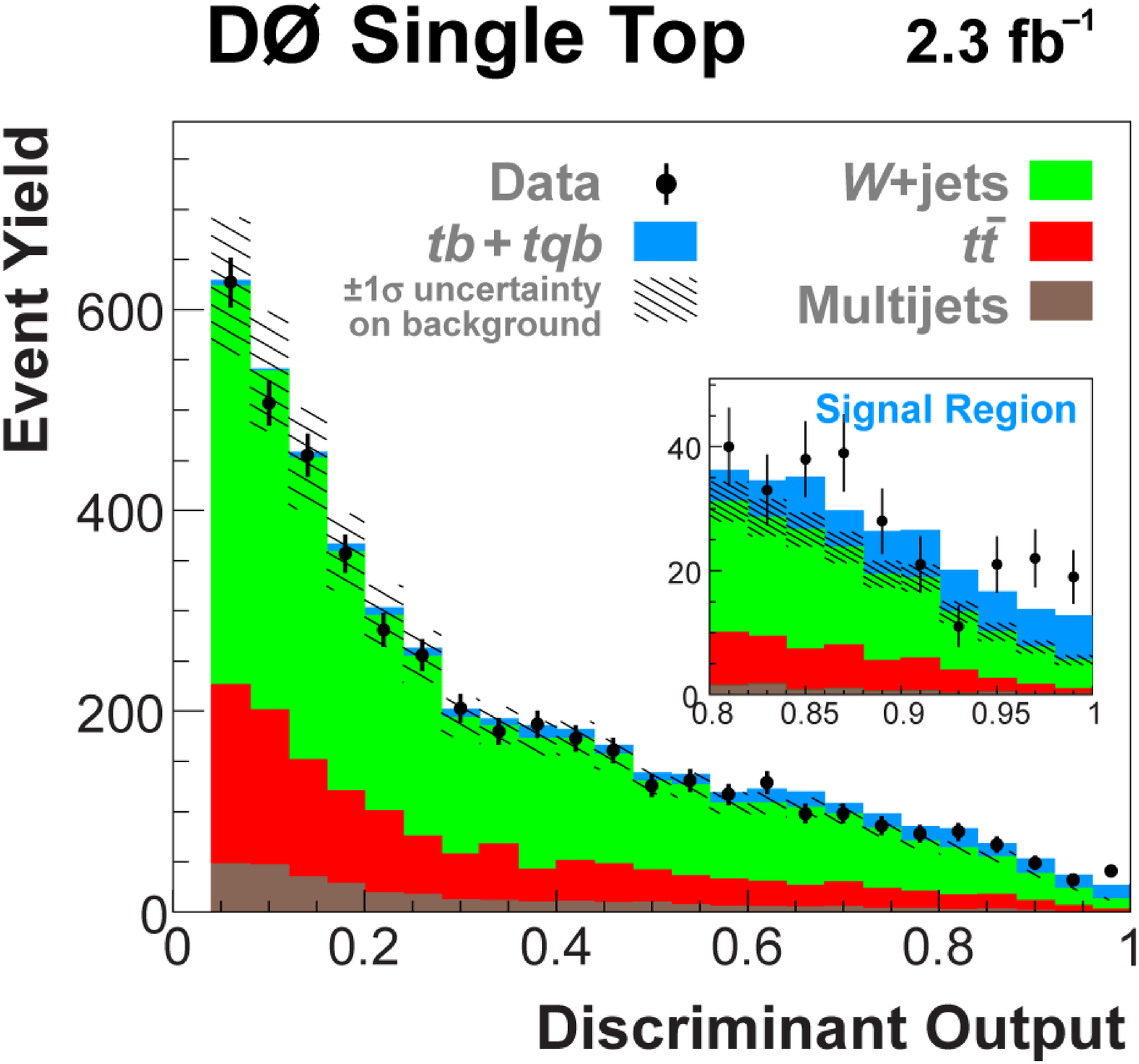,width=1.01\textwidth}
\end{minipage} 
\caption{Left (Right): CDF (D0) discriminant output distribution of the combined analysis ($\ell$+\protect\MET+jets analyses only in the CDF
case).
\label{fig:combOut}}
\end{figure}

For illustrative purposes, Fig.~\ref{fig:combOut} shows the distributions of the $\ell$+\protect\MET+jets discriminants result of the 
combination for CDF (left) and D0 (right). % and are the ones used to extract the measured cross section and the 
%signal significance (note that CDF also uses the MJ discriminant for the final result).

\section{Statistical Methods}
The cross section is measured using a Bayesian binned likelihood technique~\cite{pdg} assuming a flat non-negative prior in 
the cross section and integrating over the systematic uncertainties. The measured cross section is quoted as the 
position of the peak of the posterior density distribution
%value that maximizes the posterior likelihood, 
and the shortest interval containing 68\% of the integral of the posterior is used to set the $\pm$1 sigma uncertainties. 

The significance, in CDF, is calculated as a $p$-value~\cite{pdg}, which is the probability, assuming single top 
quark production is absent, that $-2\ln Q=-2\ln\left(p({\mathrm{data}}|s+b)/p({\mathrm{data}}|b)\right)$ is less
than that observed in the data.  
%Figure~???(c) shows the distributions of $-2\ln Q$ in pseudoexperiments that assume SM single top ($S+B$) and also 
%those that assume single top production is absent ($B$), along with the value observed in data.  

D0 also measures the significance as a $p$-value but in a different way. An ensemble of pseudo-datasets without signal contribution are generated 
and the significance is defined as the fraction of these background-only pseudo-datasets with a cross section equal to or higher 
than the measured one.

In both cases, the $p$-value is then converted into a number of standard deviations using the integral of one side of a 
Gaussian function.

\section{Cross-checks}
Before investigating the sample of selected events, both collaborations check the modeling of the 
distributions of each input variable and the discriminant outputs in data control samples depleted 
of signal.  These are the $\ell$ + $b$-tagged four-jet sample, which is enriched in $t\bar{t}$ events,
and the two- and three-jet samples in which there is no $b$-tagged jet.  The latter have high 
statistics and are enriched in $W$+jets and QCD events with kinematics similar to the $b$-tagged signal 
samples. %For the MJ analysis, two control samples are used: in the first sample, the $\METVEC$ is 
%required to be aligned along one of the jets, and in the second, the events are required to fail the 
%NNQCD requirement.  
The data distributions in the control samples are described well by the models.

\section{Systematics}
All sources of systematic uncertainty are included and correlations between normalization and discriminant 
shape changes are considered. Uncertainties in the jet energy scale, $b$-tagging efficiencies,  background modeling, lepton 
identification and trigger efficiencies, the amount of initial and final state radiation, PDFs, and factorization 
and renormalization scale have been explored and incorporated in all individual 
analyses and the combination.

\section{Results}
Table~\ref{tab:results} lists the cross sections and significances for each of the component analyses and the 
combination for each collaboration. The excess of signal-like events over the expected background is interpreted 
as observation of single top production with a $p$-value of about $3.10\times 10^{-7}$ and $2.5\times 10^{-7}$ for CDF 
and D0 respectively, corres\-ponding in both cases to a signal 
significance of 5.0 standard deviations. The sensitivity in CDF is defined to be the median expected significance and is 
found to be in excess of 5.9 standard deviations. D0
defines it as the fraction of background-only pseudo-datasets that have a measured cross section equal to or larger than the 
SM predicted cross section value and the obtained value is 4.5 standard deviations. 

CDF finds a value of the combined $s$-channel and $t$-channel cross sections of $2.3^{+0.6}_{-0.5}$~pb 
assuming a top quark mass of 175 GeV/$c^2$. D0 finds a value of  $3.94 \pm 0.88$~pb assuming a top quark 
mass of 170 GeV/$c^2$.
%The dependence on the top quark mass is +0.02~pb/(GeV/$c^2$).

Since the CKM matrix element $|V_{tb}|^2$ is proportional to the cross section, its value can be directly 
measured. From the cross section measurement at $m_t=$~175 GeV/$c^2$, CDF obtains $|V_{tb}|=0.91\pm0.11 
(\mathrm{stat+syst})\pm 0.07$(theory~\cite{harris}) and a limit $|V_{tb}|>0.71$  at the 95\% C.L. D0, from the 
cross section measurement at $m_t=$~170 GeV/$c^2$, obtains $|V_{tb}f_1^L|=1.07\pm0.12(\mathrm{stat+syst+theory})$ and a 
limit of $|V_{tb}|>0.78$  at the 95\% C.L. A flat prior in $|V_{tb}|^2$ from 0 to 1 is assumed for the 95\% CL limit results.
%This is the most precise direct measurement of $|V_{tb}|$ to date.

\begin{table}[ht!]
\caption{Left: CDF results summary for the five correlated $\ell$+\protect\MET+jets analyses combined by the SD 
analysis, the SD and the MJ analysis, and the total combination.  The LFS analysis measures only the $s$-channel 
production cross section, while the other analyses measure the sum of the $s$- and $t$-channel cross sections. 
Right: D0 results summary for the three individual analysis and their combination.}
\begin{center}
\begin{tabular}{|l|ccc||ccc|} 
\hline
 & \multicolumn{3}{c||}{CDF ($m_t=$~175 GeV/$c^2$)} & \multicolumn{3}{c|}{D0 ($m_t=$~170 GeV/$c^2$)}\\
\cline{2-7}
Analysis        & Cross                    & Significance & Sensitivity &  Cross & Significance & Sensitivity\\
                & Section (pb)             & (Std. Dev.)  & (Std. Dev.) &  Section (pb) & (Std. Dev.) &  (Std. Dev.)\\ 
\cline{0-3}\cline{5-7}
LF              & 1.6$^{+0.8}_{-0.7}$      & 2.4          & 4.0     & --- & --- &  ---   \\
ME              & 2.5$^{+0.7}_{-0.6}$      & 4.3          & 4.9     & 4.30$^{+0.99}_{-1.20}$ & 4.9 & 4.1 \\
(B)NN           & 1.8$^{+0.6}_{-0.6}$      & 3.5          & 5.2     & 4.70$^{+1.18}_{-0.93}$ & $>$5.2 & 4.1 \\
BDT             & 2.1$^{+0.7}_{-0.6}$      & 3.5          & 5.2     & 3.74$^{+0.95}_{-0.79}$ & 4.6 & 4.3 \\
LFS             & 1.5$^{+0.9}_{-0.8}$      & 2.0          & 1.1     & --- & --- & ---	\\ 
\hline
SD              & 2.1$^{+0.6}_{-0.5}$      & 4.8          & $>5.9$  & --- & --- & ---\\
MJ              & 4.9$^{+2.5}_{-2.2}$      & 2.1          & 1.4     & --- & --- & ---  \\ 
\hline
Comb.        & 2.3$^{+0.6}_{-0.5}$      & 5.0          & $>5.9$  & 3.94 $\pm$ 0.88  & 5.0 & 4.5\\ 
\hline
\end{tabular}
\end{center}
\label{tab:results}
\end{table}

\section{Conclusions}
In summary, both the CDF and D0 collaborations have developed several multivariate analysis techniques to distinguish
single top signal from background events and have combined them to precisely measure the electroweak single top 
production cross section and the CKM matrix element $|V_{tb}|$. Single top production has been observed for the first 
time by both collaborations, CDF and D0, with a significance of 5.0 standard deviations. More details can be found 
here~\cite{d0Obs,cdfObs}.

%\section*{Acknowledgments}
%The results shown here represent the work of many people. I thank my CDF and D0 colleagues for their efforts to 
%carry out these challenging physics analyses. I also thank the conference organizers for a very nice week of physics. 

%Finally, I thank the colleagues of my research institution, Fermilab, for all their help.

%We thank the Fermilab staff and the technical staffs of the participating institutions for their vital contributions. 
%This work was supported by the U.S. Department of Energy and National Science Foundation; the Italian Istituto Nazionale 
%di Fisica Nucleare; the Ministry of Education, Culture, Sports, Science and Technology of Japan; the Natural Sciences 
%and Engineering Research Council of Canada; the National Science Council of the Republic of China; the Swiss National 
%Science Foundation; the A.P. Sloan Foundation; the Bundesministerium f\"ur Bildung und Forschung, Germany; the Korean 
%Science and Engineering Foundation and the Korean Research Foundation; the Science and Technology Facilities Council 
%and the Royal Society, UK; the Institut National de Physique Nucleaire et Physique des Particules/CNRS; the Russian 
%Foundation for Basic Research; the Ministerio de Ciencia e Innovaci\'{o}n, and Programa Consolider-Ingenio 2010, Spain; 
%the Slovak R\&D Agency; and the Academy of Finland. 


\begin{thebibliography}{99}
\bibitem{cpyuan} $Q\times\eta$ is
the charge of the lepton times the pseudorapidity of the jet not assigned to be the $b$ from top quark decay; see
  C.~P.~Yuan,
%``A New Method to Detect a Heavy Top Quark at the Tevatron,''
  Phys.\ Rev.\  D {\bf 41}, 42 (1990).
\end{thebibliography}
\end{document}